\documentclass[twocolumn,amsmath,amssymb,prl,superscriptaddress]{revtex4}

\usepackage{graphicx}
\usepackage{bm}
\usepackage{hyperref}
\usepackage{psfrag}

\begin{document}

\title{Response to Comment on ``X-ray Absorption Reveals Collapse of 
Single-Band Hubbard Physics in Overdoped Cuprates''}

\author{D.C.~Peets}
\email{dpeets@scphys.kyoto-u.ac.jp}
\affiliation{Department of Physics, Graduate School of Science, Kyoto
University, Kyoto, Japan 606-8502}

\author{D.G.~Hawthorn}
\affiliation{Department of Physics \& Astronomy, University of
Waterloo, 200 University Ave.~W, Waterloo, ON, Canada N2L 3G1}

\author{K.M.~Shen}
\affiliation{Department of Physics, Cornell University, Ithaca, New
York, USA 14853}

\author{G.A.~Sawatzky}
\author{Ruixing Liang}
\author{D.A.~Bonn}
\author{W.N.~Hardy}
\affiliation{Department of Physics \& Astronomy, University of
British Columbia, 6224 Agricultural Rd., Vancouver, BC, Canada
V6T 1Z1}
\affiliation{Canadian Institute for Advanced Research, Canada}

\date{\today}

\begin{abstract}

A recent article suggested that the saturation of low energy spectral 
weight observed by X-ray absorption spectroscopy in the cuprates at high 
hole doping could be explained within the single-band Hubbard model.  We 
show that this result is an artifact of inappropriate integration 
limits.

\end{abstract}
\maketitle

Recent measurements \cite{Schneider2005,Peets2009} by X-ray absorption 
spectroscopy (XAS) have shown an abrupt saturation in the oxygen $K$ 
edge low energy spectral weight at a hole doping around $x=0.22$ in 
several different overdoped cuprate systems.  It has been argued 
\cite{Peets2009} that this could not be explained within the single-band 
Hubbard approaches commonly used to model the cuprates, possibly 
associated with the inapplicability in that doping regime of the basis 
state underpinning such approaches.  Calculations have since indicated 
that the conclusions may be more general and more significant 
\cite{Wang2010}, being apparently at odds with the three-band Hubbard 
model which the single-band model is intended to approximate.  However, 
calculations have recently been presented contending that the saturation 
can be explained within a single-band model \cite{Phillips2010}.  We 
examine this assertion.

Since the $U=6t=0.75W$ (where $t$ is the nearest neighbor hopping 
integral and $W$ is the band width in a 2 dimensional square lattice) 
spectra that Phillips and Jarrell present are, as they mention, 
continuous, with no gap below the upper Hubbard band (UHB), a nontrivial 
choice of upper integration limit is required to generate the integrated 
density of states shown in the inset to their Figure 1.  The cutoffs 
used, in units of $4t$, were $\omega_c=0.25$ for $U=6t=0.75W$ at all 
dopings and $\omega_c=0.17$ for $U=8t=W$ at all dopings, or 
$\omega_c\sim2J$ ($J$ being the antiferromagnetic nearest neighbor 
exchange interaction).  These somewhat arbitrary choices of integration 
cutoff are problematic to the point of effecting the qualitative doping 
dependence presented in their inset, for two reasons.

First, these narrow integration windows fail to capture a large fraction 
of the low energy spectral weight (i.e. the weight below the upper 
Hubbard band).  Indeed, this cutoff is near the peak of one of the 
features of interest at low dopings, inherently missing around half of 
that peak's weight.  Clear evidence of the appropriateness of the 
authors' cutoff can be seen in the unphysically low integrated weights 
in their inset --- before even considering dynamical spectal weight 
transfer (DSWT), which is central to their argument, slopes at low 
doping in a Hubbard model must exceed unity per spin state to be 
physically meaningful.  DSWT should increase this slope substantially at 
low dopings, particularly for the parameters chosen, which will tend to 
strongly enhance the spectral weight transfer.  The slope in the inset 
for $U=8t=W$ is clearly well below unity, making those points unphysical 
and in disagreement with the results in their references [7] 
\cite{Harris1967} and [8] \cite{Eskes1991}.  The slope for $U=6t=0.75W$ 
is also below unity for all segments present (this curve extrapolates to 
a positive intercept, not zero as in the physical system which is an 
insulator, because the bandwidth exceeds $U$).

A paper by Meinders {\it et al} \cite{Meinders1993} provides a useful 
demonstration of the behavior expected within the Hubbard model.  
Meinders' low energy spectral weight increases strongly with doping for 
low dopings, as seen in their Figure 3.  Meinders' points must be 
reduced by a factor of two for comparison to those in 
\cite{Phillips2010}, which are per spin. This results in expected slopes 
of about $1.5-2$ for the most comparable parameters, well above unity.  
Phillips and Jarrell used a more sophisticated approach 
\cite{Hettler1998,Jarrell2001} and in two dimensions rather than one, 
but the general arguments for spectral weight transfer are 
dimension-independent.  The factor of $\sim$2 disagreement at low 
dopings between Meinders and Phillips is anomalously large, and that the 
parameters chosen by Phillips ought to result in even steeper slopes at 
low doping makes this disagreement particularly striking.  

As for the choice of an $\omega_c$, apart from the reasoning that 
integrating the low energy spectral weight (i.e. the weight below the 
UHB) requires integrating {\slshape all} of the weight below the UHB, 
comparisons to existing theory \cite{Meinders1993,Wang2010} or 
experiment \cite{Peets2009} require integrating to the gap or minimum 
below the upper Hubbard band as was done in that previous work.  
Choosing integration cutoffs consistent with experiment is particularly 
important in this case, since the purpose of Phillips' paper is to 
explain the experimental data.  The integration cutoff must be in the 
broad minimum found well above \cite{Phillips2010}'s integration cutoff.  
The minimum below the UHB should be significantly clearer in the 
$U=8t=W$ simulations, which were not shown.

Second, and more crucially, the cutoff {\slshape must} change with 
doping.  Since hole doping (removing electrons) manifestly shifts the 
chemical potential, to which all energies in \cite{Phillips2010} are 
referenced, the low energy spectral weight and UHB are shifting with 
respect to the energy axis and integration window.  This effect is 
clearly visible in Phillips' paper itself, where the UHB weakens and 
moves to significantly higher energy on doping, consistent with the 
results in Meinders' Figure 2 \cite{Meinders1993}.  Indeed, Phillips' 
UHB shifts by roughly three times the width of the integration window 
between $x=0.05$ and $x=0.20$; the peak of the low energy spectral 
weight shifts by about half that, but it should be gaining states on its 
low energy side, so this is unsurprising.

Experimental data \cite{Peets2009} clearly show the addition of weight 
on the low energy side of the lowest-energy prepeak as holes are added, 
moving the chemical potential further below the UHB.  The upper 
integration limit in this paper, located in the minimum between the low 
energy prepeak and the UHB as mentioned above, was not changed with 
doping because zero in this case corresponds not to the chemical 
potential but to the energy of removing an electron from a 1s orbital, 
and should have negligible doping dependence (this electron does not 
exist in the single-band picture).  It should be noted that the 
experimental data are broadened by about 0.5eV due to resolution and 
core hole lifetime effects, masking small features such as any 
pseudogap;  the fact that the UHB can be clearly distinguished 
experimentally from the low-energy weight, especially at low doping, is 
indicative of a correlation-based gap significantly larger than that 
used by Philips.  Meinders \cite{Meinders1993} had the benefit of a gap 
between the low energy spectral weight and the UHB, and integrated the 
former in its entirety and none of the latter.  In this paper's second 
figure, a shift in the density of states relative to the chemical 
potential may be clearly observed.

\begin{figure}[htb]
\includegraphics[width=\columnwidth]{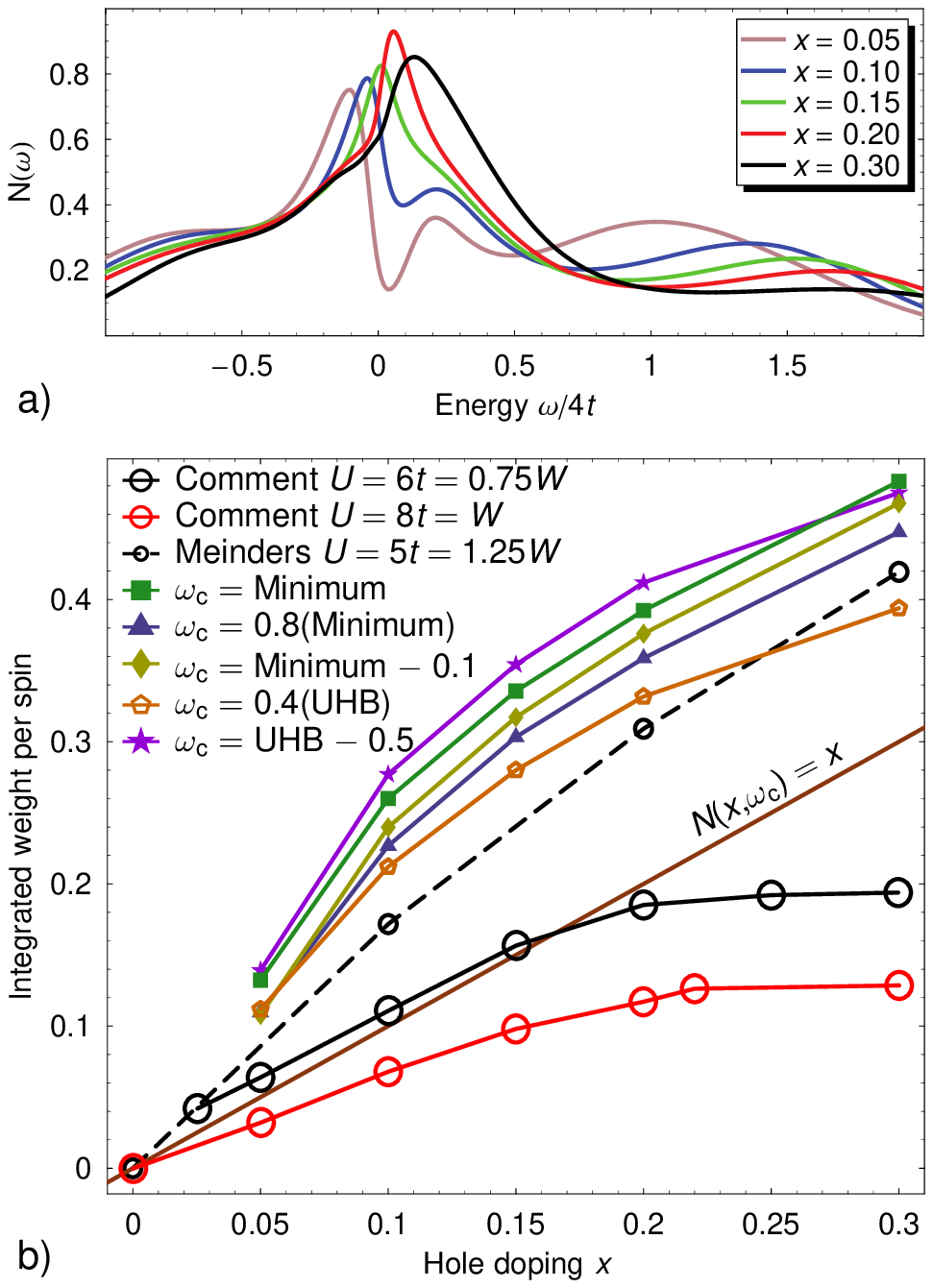}
\caption{\label{fig:inset}{\bfseries a)}Spectra from 
\cite{Phillips2010}, reproduced for reference.  {\bfseries b)}Integrated 
low energy spectral weight per spin. Points from Phillips' inset 
\cite{Phillips2010} may be compared with the results obtained when an 
upper integration limit near the minimum between the UHB and low energy 
spectral weight is chosen by five different schemes.  Curves reliant 
upon the position of the UHB are slightly suppressed at the highest 
doping due to the UHB becoming weak and indistinct.  The most directly 
comparable curve from Meinders' Figure 3 \cite{Meinders1993} has been 
included --- Phillips' calculations are for a very low $U/W$ ($W$ the 
bandwidth), and a choice of similar parameters by Meinders would have 
produced a curve above that shown. The bandwidth in 1D (Meinders) is 
$W=4t$ and in 2D (Phillips) is $W=8t$.  A line of slope 1, the minimum 
physically reasonable slope at low doping in a Hubbard model, has been 
included for reference.  The shape of the curve is not sensitive to the 
choice of cutoff near the minimum.}
\end{figure}

Failure to change the integration cutoff with doping can create 
Phillips' reported doping dependence as an artifact, since the 
integration window effectively moves to exclude more spectral weight on 
doping.  We reanalyzed Phillips' and Jarrells' spectra, integrating from 
zero to the minimum immediately below the UHB, using several different 
schemes to choose a cutoff (Figure \ref{fig:inset}).  This captured 
roughly twice as much weight, gave physically reasonable slopes at low 
doping, and consistently failed to produce saturation at high dopings.  
The qualitative results were not sensitive to reasonable choices of 
cutoffs here, so long as the cutoffs were consistent.  The reanalysed 
results have essentially the same doping evolution as that obtained by 
1D exact diagonalization \cite{Meinders1993}, and are just as 
inconsistent with the experimental data.

A further qualitative explanation for the saturation is offered in 
\cite{Phillips2010}'s final paragraph, in which the closing of the 
pseudogap halts the dynamical spectal weight transfer (DSWT).  In fact, 
this is seen in the reanalyzed weights (Figure \ref{fig:inset}), where 
the slope returns to roughly unity around $x=0.2$.  Phillips' and 
Jarrell's weights do not return to a slope of unity when the slope 
ceases to be enhanced by DSWT, they fall to a slope of nearly zero.  The 
slope per spin in a Hubbard model without DSWT is unity and with DSWT is 
well above unity, at least at low dopings, and in a Fermi liquid is 0.5.  
The Fermi liquid slope is easy to see: If we start electron doping a 
completely empty ($x=1$) Hubbard system, then each electron blocks one 
electron addition state; the UHB has vanishing weight here, so the slope 
is 1 per electron, or 0.5 per spin. If we had a $U=0$ system in this 
picture, we would start from an empty ($x=1$, not half filled) band and 
add electrons at a slope of -1 (-0.5 per spin).  Since there are two 
electron states per Cu (or one per spin), our initial value would be two 
(one) and we would cross half filling at a weight of 1 (0.5). In a 
Hubbard system at half filling there is no low-energy weight, and 
removing one electron creates two low energy electron-addition states, 
the second of which had previously been blocked by the electron thanks 
to the on-site repulsion $U$.  This leads to a slope of 2, or 1 per 
spin. Exact diagonalization in 1D has shown a smooth evolution from one 
regime to the other, with a slope of 0.5 per spin at $x=1$.  Having a 
slope of zero would require that one of these regimes evolve into the 
other nontrivially via a saddle compatible with neither picture.

Having reanalyzed Phillips' and Jarrell's calculations using integration 
limits consistent with those used on the experimental data and in 
previous theoretical work, we have shown their results to be consistent 
with 1D exact diagonalization and inconsistent with the experimental 
data.  Unfortunately, we are forced to conclude that the substance of 
\cite{Phillips2010} is based entirely on an artifact of an inappropriate 
choice of integration cutoff.  There remains no evidence that 
single-band Hubbard approaches can explain the XAS data at high dopings.

The authors gratefully acknowledge interesting discussions with P.\ 
Phillips, M.\ Jarrell, A.\ Millis, Xin Wang, and A.\ Liebsch. This work 
was supported by NSERC, the CRC program and BCSI.

{\slshape Note}: in the original PRL \cite{Peets2009}, the upper 
integration cutoff for Tl-2201 was incorrectly reported as 529.4eV, the 
same as for LSCO.  In fact, the upper integration cutoff used for 
Tl-2201 was 528.8eV.  All other cutoffs are correct in the published 
paper.  This would not have come to our attention were it not for the 
work of P.\ Phillips and M.\ Jarrell, to whom the authors are grateful.

\bibliographystyle{h-physrev}
\bibliography{commentoncomment}

\end{document}